\documentclass{moriond}

\def\be{\begin{equation}}
\def\ee{\end{equation}}
\def\bea{\begin{eqnarray}}
\def\eea{\end{eqnarray}}

\newcommand{\Photo}{\includegraphics[height=35mm]{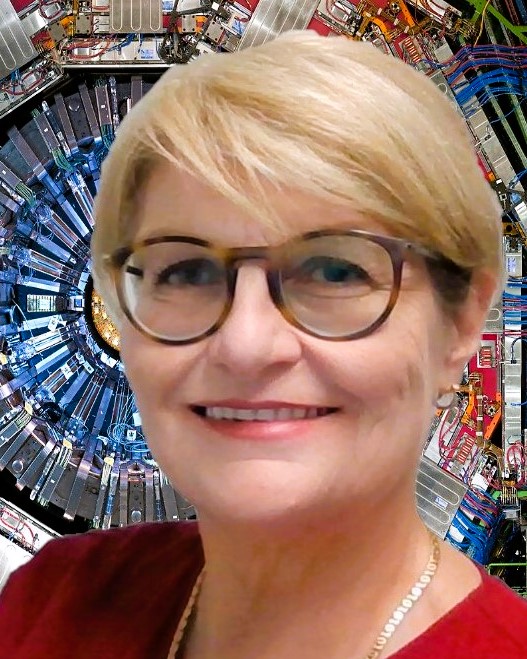}}

\begin{document}
\vspace*{4cm}
\title{Search for dark matter particle production at the LHC}

\author{ Marta Felcini~\footnote{On behalf of the ATLAS and CMS Collaborations}}

\address{University College Dublin, School of Physics,\\
Dublin, Ireland}

\maketitle\abstracts{
Understanding the fundamental nature and properties of dark matter is a main goal of fundamental physics experiments. The LHC experiments seek to detect processes that could explain how dark matter is produced and how it interacts with ordinary matter. After a reminder of the main dark matter production models, we give an overview of LHC searches for dark matter production processes and outline future opportunities. 
}

\section{Introduction}
The proof of the existence of dark matter in the universe is based on many astrophysical and cosmological measurements. While it is ascertained that dark matter interact gravitationally,  the nature of dark matter at the fundamental level remains a mystery. If general relativity is correct at all scales, as it is confirmed by available data so far,  then dark matter at the fundamental level could be composed of exotic elementary particles or composite states thereof.  
Dark matter candidate particles must have certain characteristics in the framework of the current standard model of cosmology, a six parameter model called 
$\Lambda$CDM, which so far fits precisely all available data from astrophysical and cosmological measurements. In the framework of the $\Lambda$CDM, 
dark matter constitute 26.8\% of the energy-matter content and 85\% of the  matter content of the universe.
According to the data, dark matter is largely nonbaryonic (baryonic matter is made of protons, neutrons, atoms), has the ability  to cool to non-relativistic velocities, in order to contribute to the formation of large-scale structure (such as galaxies) in the early universe, and interacts only weakly with ordinary matter, other than through gravity.

Several dark matter (DM) candidate particles and production models have been proposed with very diverse characteristics and phenomenology.  The DM candidate particle masses can go from a fraction of an eV to the TeV scale. Other properties, such as couplings of DM particles to the SM ones, which determine, among other parameters,  the production cross sections  of these new particles, are also expected to span several orders of magnitude.  For these particles and related production processes to be detectable at the LHC, with proton-proton (pp) collisions currently at a center of mass energy of 13.6 TeV,  the production cross sections have to be sizable and sufficient to produce a detectable signal, depending on LHC delivered luminosity and detection efficiency. 

The LHC experiments collect data from the high energy LHC pp collisions and analyze them to search for a vast number of processes that could produce DM candidate particles.  The experimental results are confronted to predictions by models of interactions between hypothetical dark sector particles, including DM particle candidates,  and the known standard model (SM) particles, see Fig.~\ref{fig:two}. The experiments seek to detect and measure processes leading to production of DM candidate particles in a large variety of final states. If these processes would be detected, measuring their cross sections and other physical observables, such as masses and couplings of the new particles,  would provide unique and detailed information about the nature and properties of DM candidate particles and interactions at a fundamental level. The particle physics of dark matter production studied by high energy collider experiments, in combination with results from direct and indirect DM detection experiments,  will provide a complete understanding of the nature and properties of dark matter in the universe. 

\section{Models of DM particle production at the LHC}
\begin{figure}
\begin{minipage}{0.95\linewidth}
\centerline{\includegraphics[width=0.99\linewidth]{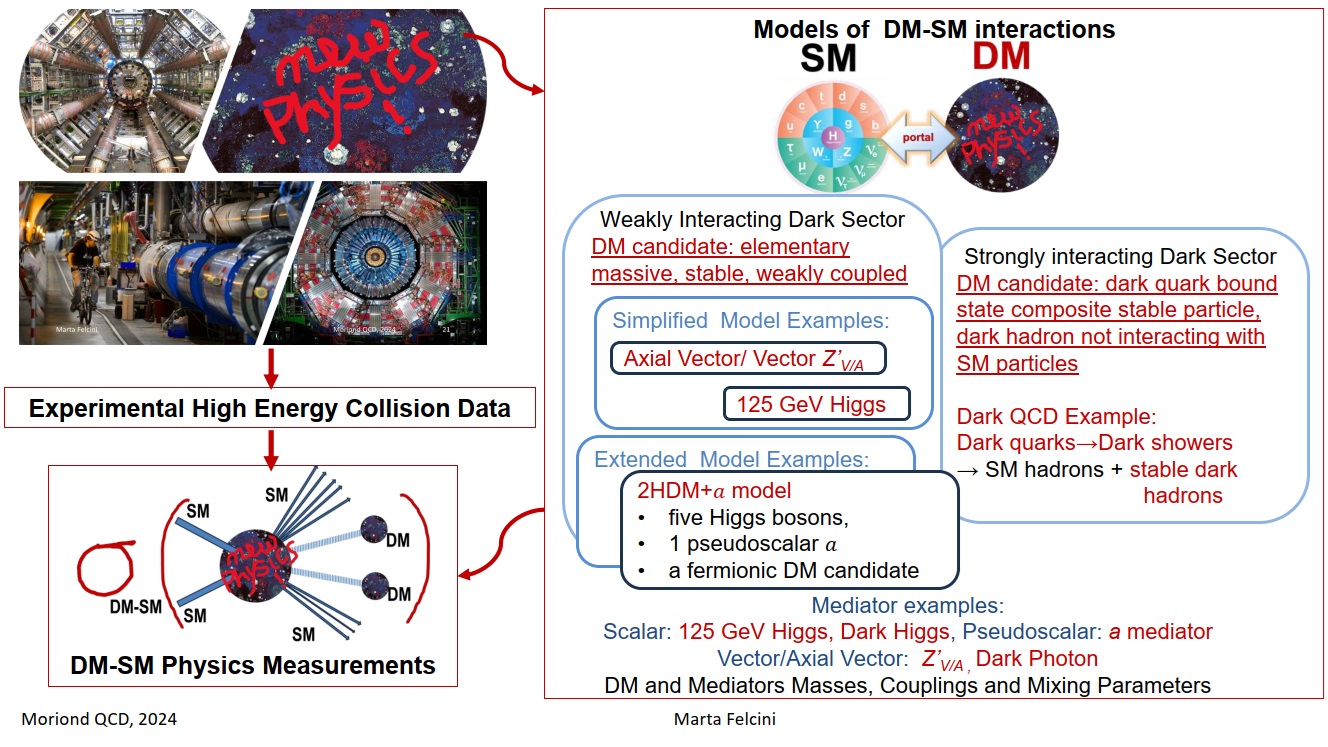}}
\end{minipage}
\caption[]{Interplay between experimental searches, models describing interactions between hypothetical dark sector particles and SM particles, and search results. Only few model examples, covered in this note, are shown.}
\label{fig:two}
\end{figure}

DM particle physics models for the LHC aim to provide a description of the interactions between particles from the SM sector and a hypothetical dark matter sector containing DM candidate particles  and interactions, mirroring the SM of known particles but in a dark version.  Many models exist describing interactions between the particles of a dark sector and the SM particles. 
These models are established on the idea that dark sector physics mirrors the physics of the SM, namely that both weakly and strongly interacting sectors exist. The picture  in Fig.~\ref{fig:two} illustrates the composition of a dark sector, in terms of weakly and strongly interacting dark sectors. There exist so called simplified models and extended models, depending on the particle contents and the phenomenology complexity. Simplified models entail few parameters, in terms of new particle masses and couplings, with relatively  simple final states, while extended models may involve several new particles, with diverse masses and couplings, and a large variety of final states and experimental signatures.   

The weakly interacting dark sector involves the existence of dark replicas of the electroweak gauge bosons, dark photon and dark Z, also known as (a version of a) Z’, as well as extensions of the Higgs sector,  including additional scalars or pseudo-scalars and/or dark replicas of the Higgs boson, called dark Higgs. An important class of models with an extended electroweak sector and suitable DM candidate particles is the case of supersymmetric models. SUSY searches at the LHC are reviewed at this conference by B.M. Dit Latour. 
In the case of extended models, other than SUSY ones, several new mediators,  which mediate the interaction between the dark sector and the SM sector, can exist, including vector and axial vector mediator such as a Z’ or a dark photon,  or scalar mediators such as the SM-like 125 GeV Higgs boson, or a dark Higgs, or pseudo-scalars  from an extended Higgs sector.  Within these models, many final states are expected, depending in particular whether the particles resulting from the DM-SM interactions decay promptly or have a delayed decay, so called long lived particles. Searches for long-lived particles at the LHC are reviewed at this conference by C. Collard.  
We will review here examples of promptly decaying particle signatures. 

The case of a strongly interacting dark sector, with the existence of dark quarks, replicas of SM quarks in the dark sector,  would lead to quite unusual consequences. DM candidates could be composite states of dark quarks, such as dark hadrons, rather than elementary particles. The final states expected for dark QCD interactions would also lead to unusual signatures such as dark showers, requiring specially designed new search techniques. 
A sample of results from experimental LHC searches in the framework of weakly and strongly interacting dark sectors is given in the next section. 

\section{Experimental searches}
\begin{figure}[t]
\begin{minipage}{0.95\linewidth}
\centerline{\includegraphics[width=0.95\linewidth]{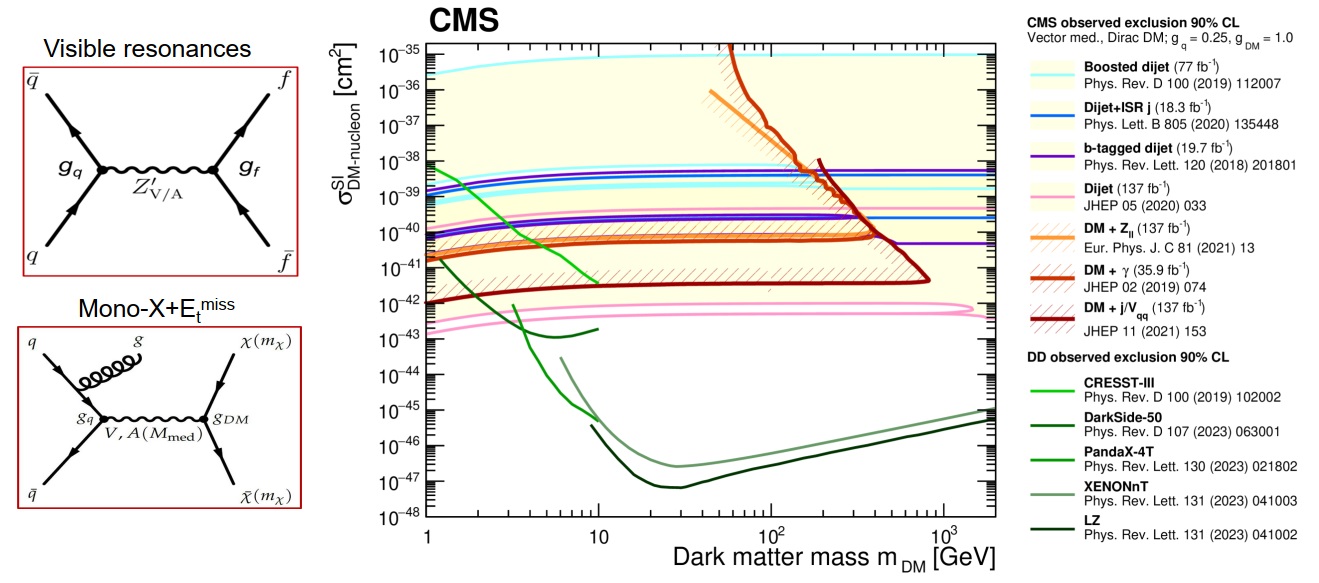}}
\end{minipage}
\caption[]{Weakly interacting dark sector: example of results for simplified models involving the exchange of a DM-SM mediator, decaying either into a visible resonance or into invisible DM particle (examples of Feynman digrams on the left-hand side),  interpreted in terms of upper limits on the DM nucleon cross section and compare to results from direct DM experiments (right hand side). Results from Ref.~\cite{CMS:2024zqs} and references therein. }
\label{fig:thr}
\end{figure}
In this section we give a brief overview of results from searches for evidences of dark sector interactions performed by the ATLAS and the CMS Collaborations with the Run 2 data, amounting to a collected sample per experiment of about 140~$\mathrm{fb^{-1}}$ of pp collisions at a center-of-mass energy of 13.6 TeV.  The results presented hereafter are taken from recent reviews by the ATLAS~\cite{ATLAS:2024fdw,ATLAS:2024lda} and CMS~\cite{CMS:2024zqs,CMS:2024zhe} Collaborations. 
Many final states are searched for in the framework of a large variety of models. A general feature of DM particle production, irrespective of the model   considered, is that the DM candidate particles are typically invisible to the detectors, so their presence leads to signatures with missing momentum in the event. Other specific features depend on the specific  model. 
In the framework of a weakly interacting dark sector, simplified models include vector or axial vector mediators with sizable couplings to both SM particles and DM particles. The mediators may then decay either into visible SM particles or into invisible DM particles.  A special case of a simplified model is the one where the SM-like 125 GeV Higgs boson decays with a small but non-negligible branching ratio into DM candidate particles.
Examples of searches in the framework of extended models are those for an extended Higgs sector with an additional pseudo-scalar mediator coupling to DM invisible particles as well as to SM particles.  We also show a representative result in the framework of a strongly interacting dark sector from searches for decays of dark hadrons  into semi-visible jets. 

For the case of simplified models of a weakly interacting sector, we show an example of results from searches for signals involving the exchange of a DM-SM mediator, a new BSM vector (or axial vector) boson (customarily indicated by a Z' symbol), produced in the collision via a Drell-Yan process (s-channel) and decaying either into visible SM fermion particles or into DM invisible particles (typical Feynman diagrams for this process are shown in the left-hand panel of Fig.~\ref{fig:thr}. In the visible case, if the mediator is produced on-shell, its decay is detectable as a visible resonance.  If the mediator decays into DM invisible particles, the event can be detected if the invisibly decaying mediator is produced in association with a visible particle, typically a photon, or a W/Z, or a jet, from initial state radiation, leading to the mono $X+p_\mathrm{T}^\mathrm{miss}$ signature. 

Several final states have been investigated as shown in the legend of  Fig.~\ref{fig:thr}. The parameters of the models are the masses of the vector boson mediators and of the DM candidate particle, as well as the coupling of the mediator to SM particles (fermions) and to DM invisible particles (mainly fermions).  The mediator can have vector or axial vector couplings, resulting  into different experimental sensitivities.  The direct outcomes of these searches, if no signal is detected, are 95\% CL upper limits on the mediator couplings to SM fermions as function of the mediator mass, as well as upper limits on the coupling of the mediator to the DM particle, as function of the DM particle mass. 
In the framework of these simplified models, with a vector or axial-vector mediator,  it is also possible to recast the experimental collider search results in terms of upper limits on the DM-nucleon scattering cross section and compare to results from direct DM search experiments. Examples of such results are shown on the left hand panel of  Fig.~\ref{fig:thr}) by the CMS Collaboration~\cite{CMS:2024zqs}. Similar results are presented by the ATLAS Collaboration~\cite{ATLAS:2024fdw}.  

\begin{figure}[t]
\begin{minipage}{0.99\linewidth}
\centerline{\includegraphics[width=0.99\linewidth]{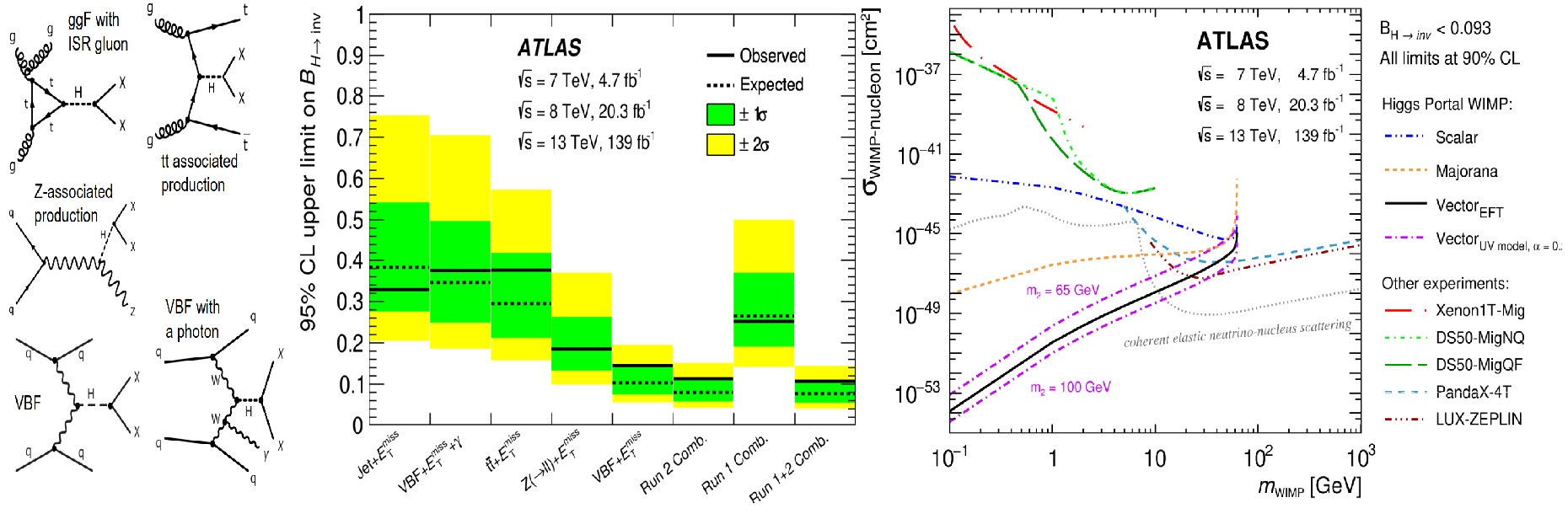}}
\end{minipage}
\caption[]{Results from searches for invisible decays of the SM-like 125 GeV Higgs boson into DM invisible particles, showing (center) the upper limits on the Higgs-to-invisible  branching fraction and (right) the spin-independent DM-nucleon scattering cross section versus DM mass, compared to direct-detection
limits.  Results from Ref.~\cite{ATLAS:2024fdw} and references therein.
}
\label{fig:fou}
\end{figure}
A special case of a simplified scalar mediator model is the one where the SM-like 125 GeV Higgs boson plays the role of the mediator between the SM and a dark sector, opening up the possibility of the 125 GeV Higgs decaying into invisible DM particles, if the DM mass is less than half the Higgs mass.  The Higgs invisible decay branching fraction would then be enhanced as compared to the expected SM value of $~\sim 0.1\%$  for $H\rightarrow ZZ^*\rightarrow 4\nu$  SM decay. The searches for invisible Higgs decays encompass many channels and final states, as suggested by examples on the left-hand panel of Fig.~\ref{fig:fou}, showing possible processes leading to invisible Higgs decay through DM particle production.   The central panel of  Fig.~\ref{fig:fou} shows the 95\% CL upper limit on the branching ratio of the Higgs boson to invisible final states in different search channels and combined, including both Run~2 and Run~1 data.  The combined upper limit on the  branching fraction of the Higgs-to-invisible final states is observed (expected)  to be 0.107 (0.077)  as reported by the ATLAS Collaboration~\cite{ATLAS:2024fdw}. This is currently the most stringent limit at the LHC from combined Run 1+2 data on the invisible Higgs decay branching fraction. The right hand panel of Fig.~\ref{fig:fou} shows the result of recasting the measured upper limit  on the 125 GeV Higgs invisible decays B$(H\rightarrow inv)$ into  upper limits on the DM-nucleon scattering cross section as function of the DM particle mass  and  comparison with  direct DM detection experiments. Comparable results with observed and expected 95\% CL upper limits of 0.15 and 0.08, respectively, are presented by the CMS Collaboration~\cite{CMS:2024zqs} and references therein.  

\begin{figure}[t]
\begin{minipage}{0.99\linewidth}
\centerline{\includegraphics[width=0.99\linewidth]{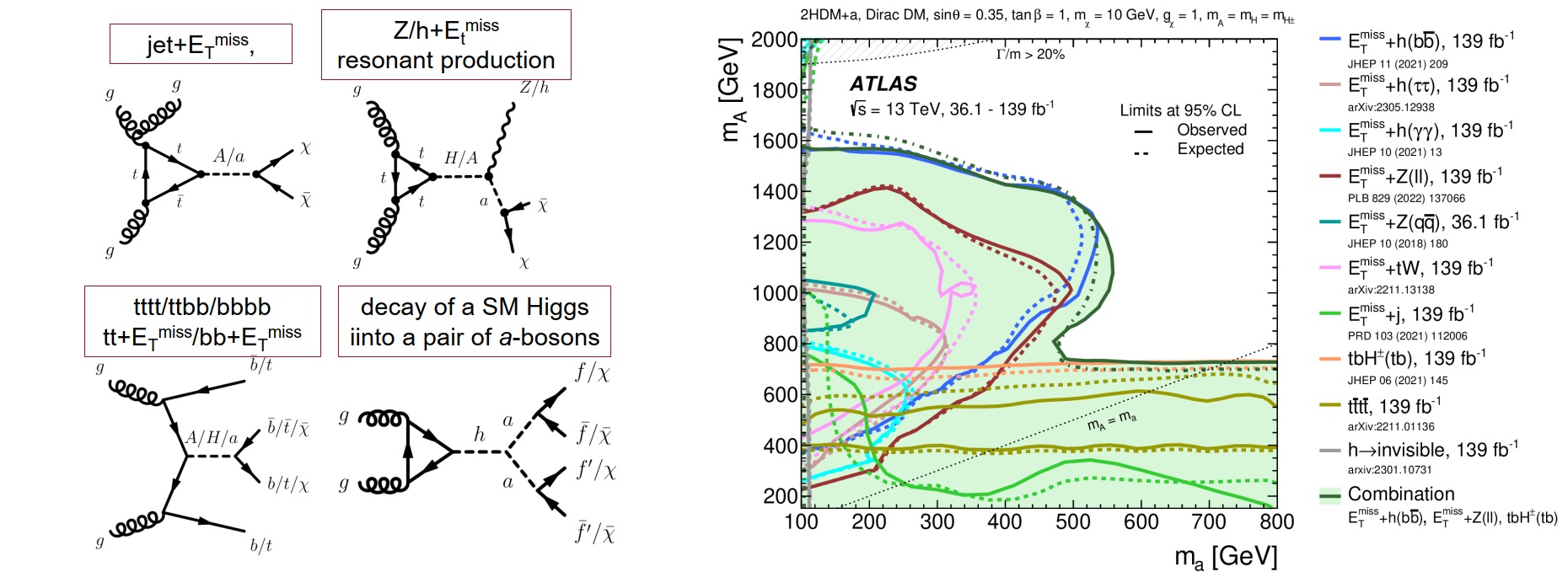}}
\end{minipage}
\caption[]{Example of results for the extended model with two Higgs doublets and a pseudo-scalar DM mediator.
Results from Ref.~\cite{ATLAS:2024fdw} and references therein.
}
\label{fig:fiv}
\end{figure}

In the framework of weakly interacting extended models,  we show the case of a two Higgs doublet model with an additional pseudo-scalar mediator $a$, the so-called  2HDM+$a$ model. A large number of processes and final states have been studied by the ATLAS Collaboration~\cite{ATLAS:2024fdw}. 
Examples of Feynman diagrams leading to various final states expected in the 2HDM+$a$ model are shown in the left hand panel of  Fig.~\ref{fig:fiv}.
Model parameters for benchmark studies are reduced from 14 to 5, including the mass of the heavy Higgs bosons,  $m_A=m_H=m_{H^\pm}$, the mass of the pseudo-scalar mediator, $m_a$, the  mass of the DM particle, $m_\chi$, mixing angle $\theta$ between the two CP-odd states $a$ and $A$,  the ratio of vevs of the two  Higgs doublets, tan $\beta$. 
A statistical combination of the most sensitive channels, $h(bb)+p_\mathrm{T}^\mathrm{miss}$ , $Z(\rightarrow \ell\ell)+p_\mathrm{T}^\mathrm{miss}$ and $tbH^\pm (tb)$, was performed. 
The constraints are evaluated for some representative benchmark scenarios, in which only one or two of the five free parameters are varied at a time.  
An example is presented in the left hand panel of  Fig.~\ref{fig:fiv},  showing the complementarity of the searches performed. 

\begin{figure}[t]
\begin{minipage}{0.99\linewidth}
\centerline{\includegraphics[width=0.99\linewidth]{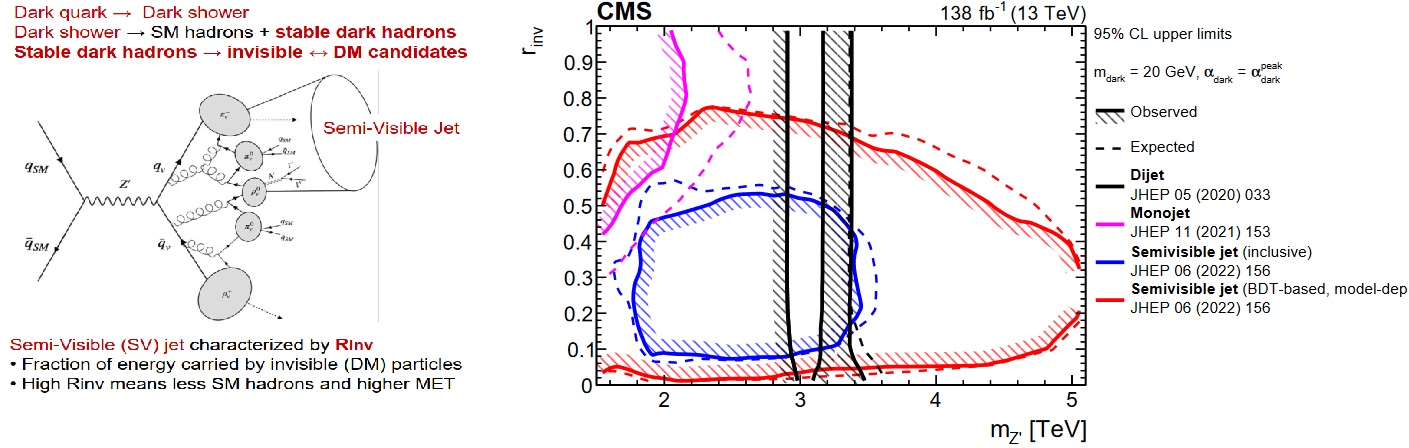}}
\end{minipage}
\caption[]{Strongly interacting dark sector search: upper limits on the mass of a Z' mediator decaying into dark quarks extends up to about 5 TeV. Ref.~\cite{CMS:2024zqs} and references therein. }
\label{fig:six}
\end{figure}

Finally we show an example of results from searches for a strongly interacting dark sector. The searches target signals of dark quark production by identifying so-called dark showers, composed of SM hadrons plus stable dark hadrons, which are neutral and non interacting with SM particles, thus invisible to the detector. Dark hadrons, composite states of dark quarks, could provide suitable DM candidates.   The presence of undetected dark hadrons in the shower leads to semi-visible jets, characterized by large values of $r_{inv}$, the ratio between the number of invisible hadrons, over the total number of (visible + invisible) hadrons, in the jet,  with a large fraction of jet energy carried away by invisible particles. An example of semi-visible jets search results is shown in  Fig.~\ref{fig:six}. Excluded regions are shown in the parameter plane  $r_{inv}$ and mass of the mediator Z', $m_{Z'}$. Limits on $m_{Z'}$ extend up to 5 TeV, depending on $r_{inv}$,  while a lower limit on  $m_{Z'}$ of about 3 TeV is set independently of $r_{inv}$ values. 
Several other searches for dark QCD and strongly interacting dark sector processes have been conducted by the ATLAS and CMS Collaborations with results documented in  Ref.~\cite{ATLAS:2024fdw} and Ref.~\cite{CMS:2024zqs}, respectively.

The experimental results show no evidence so far of weakly nor strongly interacting dark sectors. The searches will advance in the next year and decade with additional data, massive detector upgrades and new advanced technologies for data treatment in the online trigger system and the offline data analysis.  

\section{Summary}

We have given an overview of searches for dark matter particles and interactions performed by the ATLAS and CMS experiments in the LHC Run 2 data. No signal has been detected so far.  
LHC Run 3 will more than double the available LHC dataset. The High Luminosity LHC phase will extend the current statistics by an order of magnitude. The increased data statistics will largely extend  the reach of the searches for physics beyond the SM, together with detector upgrades and on line or offline data analysis techniques, which will substantially enhance the detection efficiency of the searches over larger regions of the models parameter space. 

New physics searches, and particularly the search for dark matter candidates, at the LHC, together with non collider experiments results and astrophysical and cosmological measurements, from running and planned new observatories,  have the potential to largely advance our fundamental understanding of the dark matter puzzle. 


\section*{Acknowledgments}

The author wishes to thank the Moriond QCD organizers for the very interesting conference and for the invitation to give this talk, as well as the ATLAS and CMS Collaborations for the opportunity to present recent results on their behalf.

\end{document}